\documentclass[preprint]{vgtc}               

\ifpdf
  \pdfoutput=1\relax                   
  \usepackage{graphicx}                
  \DeclareGraphicsExtensions{.pdf,.png,.jpg,.jpeg} 
\else
  \usepackage{graphicx}                
  \DeclareGraphicsExtensions{.eps}     
\fi%

\graphicspath{{figures/}{pictures/}{images/}{./}} 

\usepackage{microtype}                 
\PassOptionsToPackage{warn}{textcomp}  
\usepackage{textcomp}                  
\usepackage{mathptmx}                  
\usepackage{times}                     
\usepackage{cite}                      
\usepackage{tabu}                      
\usepackage{booktabs}                  
\usepackage{threeparttable}
\usepackage{verbatim}
\usepackage{enumitem}
\usepackage{soul}
\usepackage{hyperref}

\newcommand{\emily}[1]{\textcolor{magenta}{#1}}
\newcommand{\kristin}[1]{\textcolor{teal}{#1}}

\newcommand{\remove}[1]{}

\newcommand{\revise}[1]{\textcolor{black}{#1}}

\onlineid{0}

\vgtccategory{Research}

\vgtcinsertpkg

\title{Let's Get Vysical: Perceptual Accuracy in Visual \& Tactile Encodings}

\author{Zhongzheng Xu 
\thanks{e-mail: zhongzheng.xu@emory.edu}\\ %
        \scriptsize Emory University %
\and Kristin Williams\thanks{e-mail: kristin.williams@emory.edu}\\ %
     \scriptsize Emory University %
\and Emily Wall\thanks{e-mail: emily.wall@emory.edu}\\ %
    \scriptsize Emory University %
}

\abstract{
In this paper, we explore the effectiveness of tactile data encodings using swell paper in comparison to visual encodings displayed with SVGs for data perception tasks. By replicating and adapting Cleveland and McGill's graphical perception study for the tactile modality, we establish a novel tactile encoding hierarchy. In a study with 12 university students, we found that participants perceived visual encodings more accurately when comparing values, judging their ratios with lower cognitive load, and better self-evaluated performance than tactile encodings. However, tactile encodings differed from their visual counterparts in terms of how accurately values could be decoded from them. This suggests that data physicalizations will require different design guidance than that developed for visual encodings. By providing empirical evidence for the perceptual accuracy of tactile encodings, our work contributes to foundational research on forms of data representation that prioritize tactile perception such as tactile graphics.
}

\CCScatlist{
  \CCScatTwelve{Human-centered computing}{Visu\-al\-iza\-tion}{Visu\-al\-iza\-tion techniques}{};
  \CCScatTwelve{Human-centered computing}{Visu\-al\-iza\-tion}{Visualization design and evaluation methods}{}
}

\begin{document}


\firstsection{Introduction}

\maketitle


\revise{\begin{figure}[t]
  \centering
  \includegraphics[width=0.95\linewidth]{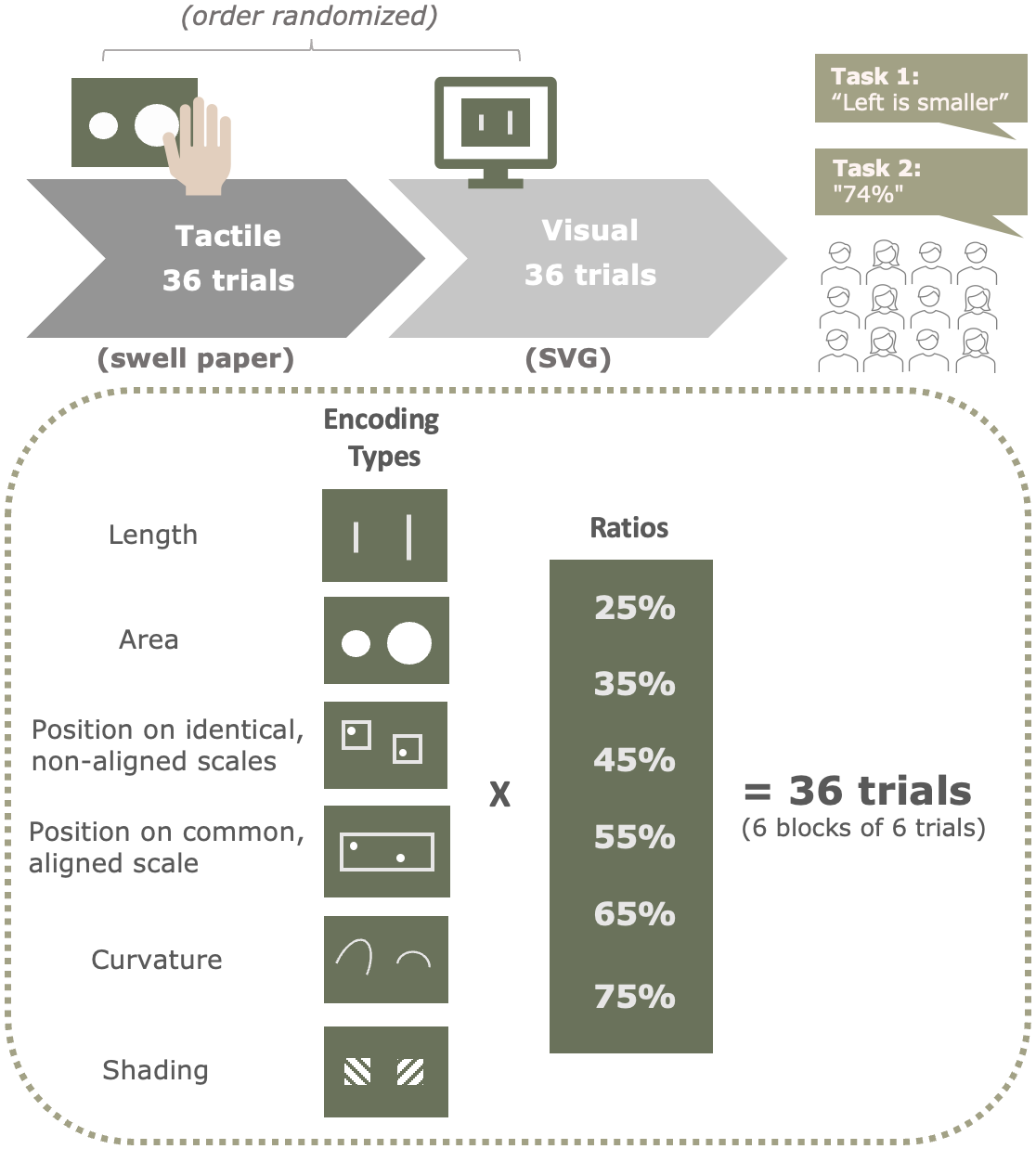}
  \caption{Summary of experiment design.
  }
  \vspace{-1.5em}
  \label{fig:design}
\end{figure}}
 
\vspace{-1em}
Since Data Visualization emerged as a research field, scientists have incorporated findings from visual perception and cognition to develop guidance on effective and efficient ways of communicating data \cite{cleveland1984graphical, healey2007perception}. For example, experiments on \textit{preattentive features} shed light on basic visual properties that can be readily detected by low-level visual systems \cite{healey2007perception, treisman1985preattentive, treisman1992automaticity}. Such results inform design guidelines\cite{fekete2008value}, and have enabled both researchers and designers to leverage empirical findings to develop effective visualizations \cite{ward2010interactive}. 
The efficacy of bar charts, for instance, is supported by empirical findings that humans are best at judging positions and lengths, compared to other encodings such as area or curvature\cite{cleveland1986experiment, cleveland1984graphical}. 
In fact, there is a wealth of prior work on graphical perception, dating back several decades and continuing today, grounding guidance on visual representations of data in empirical studies \cite{davis2022risks, mccoleman2021rethinking, treisman1985preattentive, cleveland1984graphical, cleveland1986experiment}. 

While tangible representations of data pre-date the field of data visualization, data physicalization has garnered recent interest, building upon momentum in data visualization, tangible user interfaces, and data accessibility \cite{jansen2015opportunities, shaer2010tangible, edman1992tactile}.
Data physicalization enables the creation of tangible representations of data that can be perceived and interpreted through tactile senses. 
By leveraging haptic perception, data physicalization can enhance user experience and engagement with data. 
For example, a tactile map representing geographic data can provide a unique and interactive experience by enabling touch-based navigation and exploration of the data \cite{butler2021technology, holloway20223d, taylor16}.
Moreover, data physicalization provides a medium through which individuals with visual or auditory impairments can access information \cite{lundgard2019sociotechnical}.

For data physicalization to be effective, however, we need a better understanding of the unique affordances and strengths each perceptual modality allows through different encoding types.
To date, visualization researchers acknowledge that we have yet to move beyond creating 3D versions of familiar visual graphics even when creating them for accessibility \cite{jansen2015opportunities, lundgard2019sociotechnical}.
This is problematic, as designing to privilege tactile senses may require fundamentally different considerations than designing to privilege visual senses, research that currently lacks empirical support \cite{jansen2015opportunities}. 
For instance, it must account for differences in tactile acuity, known to decline with age yet retaining high sensitivity in individuals who are literate in Braille \cite{wolfe2006sensation}. 
This underscores the need for foundational research in tactile perception for data physicalization. 
\textit{To address the current gap in empirical support for data physicalization design}, we conducted a within-subjects study comparing tactile and visual perception of six encoding types (position aligned, position non-aligned, length, area, shading, and curvature) with 12 university students. 
We find that the same order of perceptual accuracy of encoding types found by Cleveland and McGill~\cite{cleveland1984graphical,cleveland1986experiment} does not hold for tactile graphics. 


\section{Related Work}
Below, we first describe Cleveland and McGill's graphical perception study (1986) and its subsequent influence on the field of data visualization before covering tactile perception and data physicalization. \revise{Cleveland and McGill's study laid the groundwork for data visualization by identifying elementary perceptual tasks to inform graphical design, subsequently inspiring studies that expanded on their findings. As the field of data physicalization, particularly tactile graphics, gains momentum with innovative methods, there is a need for comprehensive and systematic research on tactile data encodings. }

\subsection{Graphical Perception Studies}
Cleveland and McGill identified elementary perceptual tasks for data visualization to inform graphical design. In their 1984 study, Cleveland and McGill investigated the efficiency and accuracy of decoding numerical values from visual, graphical representations \cite{cleveland1984graphical}. Inspired by other scientific fields and Julesz's concept of \textquotedblleft pre-attentive particles", they established axioms of data visualization \cite{julesz1981theory,cleveland1984graphical}. To do so, they identified 10 elementary perceptual tasks they thought were employed either individually or multiply when analyzing visual data representations such as stacked bar charts and area plots \cite{cleveland1984graphical}. In a 1986 follow-up experiment, they presented participants with isolated stimuli designed around these 10 tasks to eliminate context or potential confounding factors to uncover the perceptual accuracy for each encoding type \cite{cleveland1986experiment}. They then proposed a hierarchy of visual encoding types in order from most to least accurate: position judgments both 1) along a common scale and 2) non-aligned identical scales, 3) length, 4) direction 5) angle, 6) slope, 7) area, 8) volume, 9) shading, and 10) saturation judgments.
Cleveland and McGill's systematic taxonomy of graphical perception had clear implications for designing data visualization.

Cleveland and McGill's work sparked a multitude of follow-up studies in data visualization.
By distilling target visual elements as \textquotedblleft visual variables", they could be used to investigate accuracy and response time in visual tasks, show how to identify other fundamental visualization elements like size, shape, and color, or to compare different chart types \cite{simkin1987information, spence1991displaying, tremmel1995visual, ware2012information, healey2012attention, rosenholtz2005measuring, wigdor2007perception, stone2008alpha, mccoleman2021rethinking}.
\revise{More recently, Heer and Bostock replicated Cleveland and McGill's findings~\cite{cleveland1984graphical} to demonstrate the viability of crowdsourcing platforms for graphical perception experiments~\cite{heer2010crowdsourcing}.} Researchers have extended Cleveland and McGill's elementary tasks to examine how visual variables interact and combine to uncover capacity limits in attention or perception \cite{haroz2012capacity, ware2019information, cui2021synergy, quadri2021survey}.

\subsection{Data Physicalization and Tactile Perception}
Data physicalization\textemdash a term recently created by researchers\textemdash is the process of representing data through physical objects \cite{jansen2015opportunities}. A recent focus in this field has been finding alternatives to overcome cost and reproducibility limitations \cite{brittell2018usability, giraud2017map, gual2015improving}. For example, Brittel compared the impact of fabrication materials like 3D printed plastic and microcapsule paper on the discriminability of map symbols \cite{brittell2018usability}. In addition to fabrication techniques, novel approaches such as actuated displays, microrobots integrated into tablets and multimodal methods combining haptics and audio are being explored \cite{guinness2019robographics, giudice2012learning, siu2018shapeshift, stangl20143d}. 

A systematic review of data physicalization from 2010 to 2020 shows that tactile graphics remain the most commonly used production technique in the decade\cite{butler2021technology}. When designing chart types for data physicalization, we continue to heavily borrow established chart types from the field of visualization, such as bar plots or line charts \cite{hahn2019comprehension, butler2021technology, guinness2019robographics}. There is a critical need for empirical evidence supporting how physical variables\textemdash a tactile equivalent of \textit{visual variables}\textemdash affect the efficiency of perceiving tactile graphs, similar to the way Cleveland and McGill's contribution to data visualization \cite{jansen2015opportunities}.

The sense of touch, or tactile perception, has been studied in various fields from haptics to psychology and neuroscience. The tactile modality supports perception of textures, shapes, and objects \cite{lederman1987hand}. Lederman and Klatzky proposed a tactile encoding hierarchy, focusing on unique physical aspects like vibration, temperature, and roughness \cite{lederman1987hand}. It is important to note that Lederman and Klatzky's ranking explores different properties compared to our study, as we focused on elementary visual encodings and their comparable tactile representations.

\section{Study Design}
We used a within-subjects design to compare perceptual accuracy across visual and tactile perception using support vector graphics (SVGs) rendered in a web browser (visual) and printed on swell paper with a Thermoform printer (tactile).  \revise{We also incorporated a tabular condition as sanity check (analysis of which is included in Supplemental Materials)}. A key objective was to explore to what extent prior findings on the accuracy of encoding types in graphical perception hold for tactile perception.
The study design is summarized in Figure~\ref{fig:design}.


\medskip
\noindent\textbf{Participants.}
We recruited 12 (5 female and 7 male) undergraduate college students through student listservs at a US University. Studies took place in a lab setting 
and were designed to last 1-1.5 hours. 

\medskip
\noindent\textbf{Comparing Tactile and Visual Perception.}
We employed a within-subjects study design with perceptual condition (visual, tactile\remove{, tabular}) and encoding type (position aligned, position non-aligned, length, area, shading, and curvature) as our independent variables. 
Their order was randomized and counterbalanced across participants. Participants completed 36 trials for each condition, during which they sat next to the experimenter so that the experimenter could advance to the next trial and manually record their responses. To enable equivalent data collection in both the visual and tactile conditions, participants were asked to provide their answers verbally.

\medskip
\noindent\textbf{Encoding Type.}
Within each perceptual condition, we investigated differences when judging proportional differences (ratios) in stimuli across six encoding types: position aligned, position non-aligned, length, area, shading, and
curvature. These were from the original Cleveland and McGill study (see the ``Encoding Types" column in Figure \ref{fig:design}) \cite{cleveland1984graphical, cleveland1986experiment}.
It identified 6 ranks for 10 elementary perceptual tasks, and so, we chose only one from each of the tasks tied in rank as follows. 
\revise{We selected length from \{length, direction, angle\} as it is used in one of the most common chart types in early education (bar charts)~\cite{alper2017visualization}; curvature from \{volume, curvature\} and shading from \{shading, color saturation\} they can be more easily compared across the conditions without favoring one over the other (furthermore, volume and color saturation cannot be encoded using our chosen tactile modality).} To collect multiple responses per encoding type we tested ratios ranging from 25-75\% in intervals of 10 so that each encoding type was tested in 6 trials. For each trial, we asked participants 2 questions in turn:\vspace{-0.5em}
\begin{enumerate}[noitemsep]
    \item Identify the smaller of each stimuli pair (Forced Choice) 
    \item Estimate the proportion (between $0-100$) of the smaller stimulus relative to the larger (Estimation) 
\end{enumerate}

\medskip
\noindent\textbf{Trial Blocks}
Trials within the tactile and visual conditions were completed as 6 blocks of 6 trials each for a total of 36 unique trials in each condition. 
Within each block, each encoding type appeared exactly once, as did each ratio.
The order of the trials within each block and the order of the blocks was randomized. 
See Figure \ref{fig:design}.

\medskip
\noindent\textbf{Apparatus.}
To compare across visual and tactile perception, we created a set of 36 SVG files for each of the 6 encoding types and 6 ratios (in equal 10\% intervals from 25\% to 75\% using our arbitrarily chosen base standard value). 
Participants used the Microsoft Edge web browser on a Lenovo ThinkPad T14 Gen 3 with a 14" display for the visual condition. We printed the SVG files on Swell Touch paper\textemdash a microcapsule paper\textemdash and used the \textit{American Thermoform} machine to raise their graphics  for the tactile condition \cite{americanthermoform}. Participants were blindfolded for this condition. Example images of the visual and tactile conditions are displayed in Figure \ref{fig: stimuli}. 
For each trial, participants were presented with a pair of stimuli of the given type. In the visual condition, page adornments were limited to two buttons suitable for navigating within the trial block. In the tactile condition, each pair was printed on the Swell Form paper without any adornments. \revise{We have uploaded the files and code used in our study to our Github repository, which can be accessed via the link: \href{https://github.com/CAV-Lab/vis_tactile_vis23_supplemental}{https://github.com/CAV-Lab/vis\textunderscore tactile\textunderscore vis23\textunderscore supplemental}.} 

\remove{
\begin{figure}[t]
  \centering
  \includegraphics[width=0.95\linewidth]{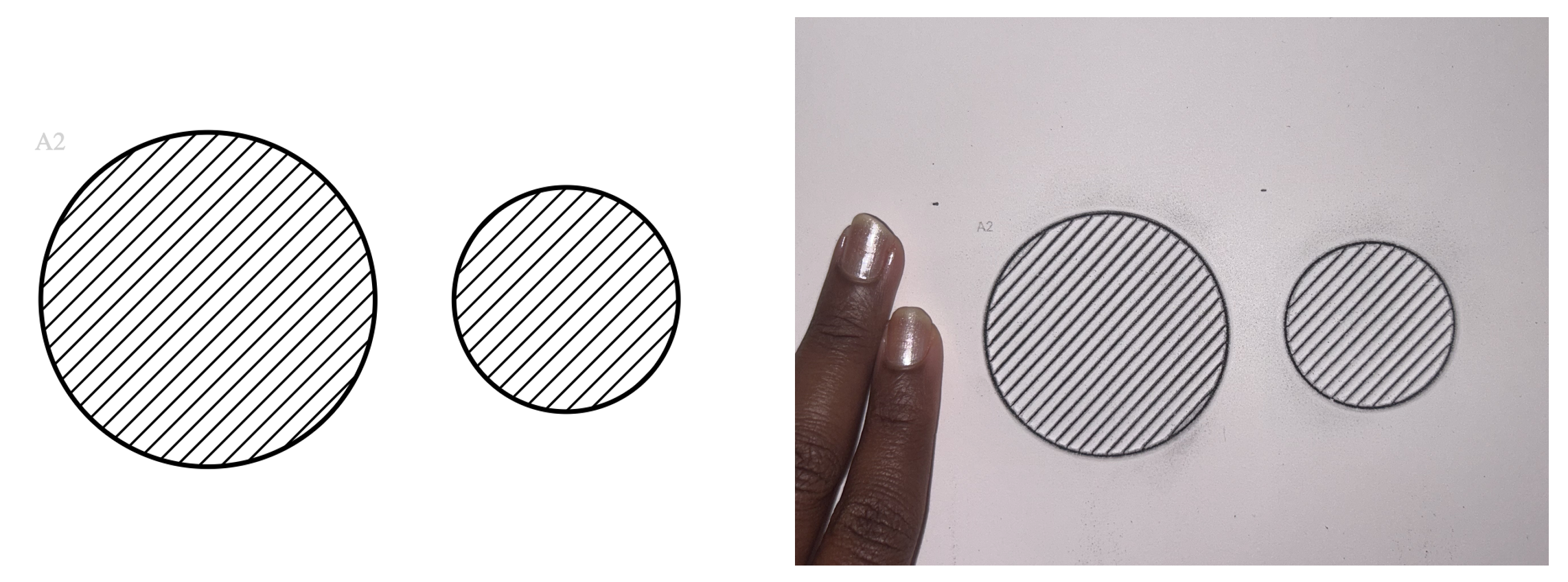}
  \caption{Example trial images for the \textit{Area} encoding type comparing the screen-based visual condition (left) and swell paper tactile condition (right).
  }
  \vspace{-2em}
  \label{fig: stimuli}
\end{figure}
}

\medskip
\subsection{Procedure}
The study lasted up to 1.5 hours and was video recorded. 
After providing informed consent, participants completed 4 main activities: a survey, condition 1, condition 2, \remove{modality 3,} and a semi-structured interview. 
\remove{Visual, tactile, and tabular}
First, participants completed a survey covering basic demographic questions and their background and experience with visual and tactile data representations. Next, they completed each of 2 conditions (tactile, visual) in a randomly assigned order. Each condition began with 6 practice tasks. During the practice session, participants were introduced to each encoding type in the given modality and the set of questions they would be asked to answer. During this time, the experimenter answered any questions participants had regarding the encoding types or the task instructions. 
Once participants completed all of the practice trials and indicated their comfort with the tasks, they proceeded to complete the 6 blocks of trials for that condition. Following the completion of each condition, participants rated their perceived accuracy on a 7-point Likert scale (1 = least accurate, 7 = most accurate) and cognitive load on a 10-point scale using a subset of the NASA TLX questionnaire\cite{hart2006nasa}. Then, participants completed the second condition followed by their ratings. Lastly, participants completed a semi-structured interview covering their experience with the trials, task strategies, and preferences for encoding types.

\subsection{Hypotheses}
Based on prior research, we hypothesize that we will be able to replicate prior results from Cleveland and McGill in the visual condition, obtain novel perceptual accuracy results for the tactile condition, and identify trade-offs between visual and tactile stimuli types.  

\begin{itemize}[noitemsep]
\item[] \textit{H1: Comparison Accuracy.} Participants will compare data values more accurately using visual encodings\remove{, followed by tabular encodings, and finally,} \revise{than} tactile encodings.
\item[] \textit{H2: Proportional Accuracy.} Participants will perceive \textit{proportion differences} more accurately when presented with \remove{tabular encodings, followed by}visual encodings\remove{, and lastly,} \revise{than} tactile encodings.
\item[] \textit{H3: Encoding Type Order of Accuracy.} Participants will perceive the \textit{proportional difference} between visual encodings in the order of accuracy found in Cleveland and McGill's experiments: position aligned, position non-aligned, length, area, curvature, shading \cite{cleveland1984graphical}. However, this order will differ for comparable tactile encodings. \remove{Participant accuracy for tabular encoding will surpass all visual and tactile encodings.}
\item[] \textit{H4: Cognitive Load.} Participants will experience higher cognitive load for \remove{ the tabular condition, followed by}the tactile condition\remove{, and finally,} \revise{than} the visual condition.
\item[] \textit{H5: Perceived Accuracy.} Participants will perceive their performance to be better for visual encodings \remove{, followed by table encodings, and lastly,}\revise{than} tactile encodings.
\end{itemize}

\subsection{Measures}
\begin{itemize}[noitemsep]
    \item[] \textit{H1: Comparison Accuracy.} We \revise{counted the number} of correct answers (to the question \textit{identify the smaller of each stimuli pair}) for the tactile and visual condition respectively. The count for each condition was then compared. 
    \item[] \textit{H2: Proportional Accuracy.} We adopted the same methodology in Cleveland and McGill's study \cite{cleveland1986experiment}. For each response, we computed the absolute error (AE) within each  conditions using: $AE = |Responded Proportion-True Proportion|$, where \textit{RespondedProportion} represents the participant's response for the proportion judgment, and \textit{TrueProportion} refers to the actual encoded ratio for the corresponding trial. \remove{MAE was thus calculated by averaging values across all encoding types and participants for each modality.}
    \item[] \textit{H3: Encoding Type Order of Accuracy.} We evaluated and compared the Mean Absolute Error (MAE) of each encoding type type within conditions. \remove{In addition, we employed descriptive analysis to compare the MAE of tabular condition trials.}
    \item[] \textit{H4: Cognitive Load.} We adapted the NASA TLX questions to measure participants' cognitive load during the tasks. \revise{Participants were asked to rate each modality according to the level of mental demand (``simple vs. demanding'') and frustration (``relaxed vs. stressed'') on a scale from 1 to 10 (1 = low, 10 = high).} Questions on physical demand, temporal demand, and effort were excluded since our study did not involve any physically demanding tasks, and participants were free to choose their own pace and take breaks.
    \item[] \textit{H5: Perceived Accuracy.} We asked participants to rate how accurate they thought they were when completing the tasks in each  \remove{of the three}condition using a 7-point Likert scale (1 = least accurate, 7 = most accurate).
   
\end{itemize}

\section{Results}
We found that the accuracy of perceiving proportional differences from elementary visual encodings, when presented in their tactile forms\remove{ on embossed paper,} \textit{deviates} from the established visual hierarchy.
Before running any statistical test to assess H2 and H3, we cleaned the data by removing outliers which we defined as \textgreater3 SD.

\medskip
\noindent\textbf{H1: Comparison Accuracy.} 
Participants were slightly more accurate when completing tasks using visual encodings (3/432 incorrect) than when using tactile encodings (10/432 incorrect). However, a Matched Pairs Signed test revealed the encoding type did not have a significant effect on task accuracy (\textit{M} = -3.500, \textit{p} = 0.092). \remove{Our results fail to support H1.} \revise{Our results show that there is no significant difference in accuracy when participants identify the smaller stimulus in the pair.}\

\medskip
\noindent\textbf{H2: Proportional Accuracy.} 
  The MAE for the tactile condition was 12.60\% \revise{($SD=10.21\%$, $range=0-50\%$)}, and the MAE for the visual condition was 10.25\% \revise{($SD=8.65\%$, $range=0-50\%$)}. Because the data were not normally distributed, we applied Aligned Rank Transform (ART) to the cleaned data, then performed Repeated Measures Analysis of Variance (RMANOVA) for the visual and tactile conditions, revealing a significant difference between them ($F(1, 836) = 4.651$, $p = 0.031$). In our contrast test, we used Holm-adjusted $p$-values, and the result between the tactile and visual conditions was $t(836)=2.146$, $p=0.032$. \revise{Our results show that there is a significant difference in participants' perceptual accuracy between the visual and tactile condition.} 

\begin{figure}[t]
    \centering
    \includegraphics[width=0.95\linewidth]{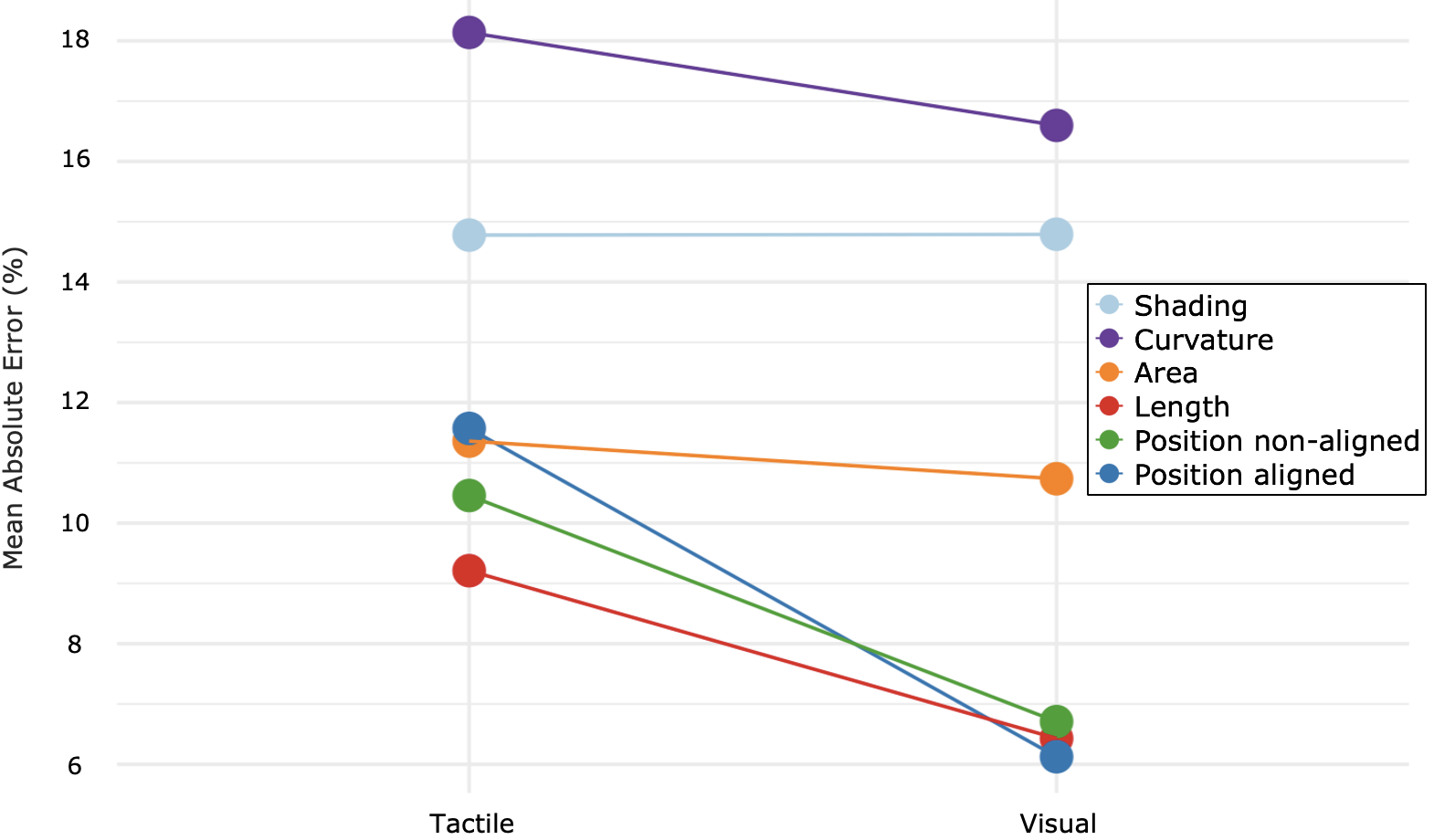}
    \caption{\revise{MAE of the six encoding types for tactile (left) and visual (right). Lower MAE corresponds to higher accuracy. 
    In the legend, we present the visual encoding hierarchy hypothesized by Cleveland and McGill \cite{cleveland1984graphical, cleveland1986experiment}. \remove{An asterisk (*) indicates that the encoding type was not examined in their experiments.}
    }
    }
    \vspace{-1.5em}
    \label{fig:rank2}
\end{figure}

\medskip
\noindent\textbf{H3: Encoding Type Order of Accuracy.} 
\remove{We performed RMANOVA to compare the MAE of encoding types within the visual ($(F(5, 55) = 13.51$, $p < 0.001$, $\eta ^ 2 = 0.49, \epsilon = 0.51$)), with a Greenhouse Geisser corrected $p < 0.001$, and tactile conditions ($(F(5, 55) = 4.57$, $p = 0.0015$, $\eta ^ 2 = 0.21, \epsilon = 0.73)$).}\revise{Analyses performed on the cleaned data using ART and RMANOVA revealed a significant main effect for encoding type ($F(5, 836) = 24.504$, $p < .001$).} Figure \ref{fig:rank2} presents the MAE for each encoding type and modality, with the legend showing Cleveland and McGill's ranking for comparison \cite{cleveland1984graphical, cleveland1986experiment}. \remove{Our findings for the visual condition partially aligns with the hierarchy established in the Cleveland and McGill paper \cite{cleveland1984graphical, cleveland1986experiment}. } 
\revise{
We observe a clear difference in the order for shading (light blue) and curvature (purple); whereas Cleveland and McGill posit curvature to be more accurate than shading~\cite{cleveland1984graphical}, we observe the opposite.
Nonetheless, these two encoding types remain amongst the lower half of encoding types by perceptual accuracy. 
At the top of the ranking, Cleveland and McGill's findings indicate that the three highest accuracy encoding types are, from most to least accurate, position aligned, position non-aligned, and length~\cite{cleveland1984graphical}; our results, while all relatively close in terms of MAE, flip the second and third most accurate encodings. 
} 

\revise{Figure \ref{fig:rank2} further demonstrates that the perceptual accuracy hierarchy for tactile encodings (left) varies compared to the visual hierarchy (right).}
In the tactile condition, ``length" emerged as the encoding with the lowest error in participants' judgments. Interestingly, ``position along a common scale" demonstrated the lowest error in the visual condition but ranked in the lower half for the tactile condition. ``Area," classified in the lower half of the visual hierarchy, ranks as the third-best encoding in the tactile hierarchy, although the MAE is only marginally different between tactile and visual. 
\revise{However, in our subsequent contrast tests using the Holm-adjustment, we compared each encoding type with all others within both visual and tactile conditions. ``Curvature" was the only encoding that showed a significant difference from five other encoding types in the visual condition and from four types in the tactile condition.
Only ``position-aligned" showcased a significant difference between the visual and tactile conditions. 
In summary, we observe notable differences in our visual hierarchy compared to that obtained by Cleveland and McGill, and, although we observed differences between tactile and visual rankings qualitatively, our findings didn't indicate any significant distinction between the tactile and visual hierarchies.}

\remove{In addition, the tabular condition had an overall MAE of 6.07\%, lower than all visual and tactile encodings. These findings confirm H3. }

\medskip
\noindent\textbf{H4: Cognitive Load.} 
Using a scale ranging from $1-10$ (1 = low, 10 = high cognitive load), we asked participants to rate each modality according to the level of \revise{mental demand (``simple vs. demanding'') and frustration (``relaxed vs. stressed'').} \revise{Participants rated tactile as requiring higher \textit{mental demand} than visual: tactile received an average score of 6.50 / 10 ($SD=2.39$, $range=2-8$) and visual received an average 5.50 / 10 ($SD=2.32$, $range=2-9$), although the difference was not statistically significant (a Wilcoxon test indicated $V=35$ and $p=0.1522$). 
Participants rated tactile as more \textit{frustrating} than visual: tactile received an average score of 4.33 / 10 ($SD=2.19$, $range=2-9$) and visual received an average 3.75 / 10 ($SD=1.60$, $range=1-7$); however, the difference was likewise not statistically significant (a Wilcoxon test indicated $V=31$ and $p=0.3346$).} \revise{While qualitative data suggests that the tactile condition leads to greater mental demand and frustration, we do not observe a significant difference in cognitive load between the two modalities.}
\remove{These results \textbf{fail to support H4}. 
We hypothesize that the low cognitive load score for the tabular condition may be due to its simplified presentation, which directly displayed two numbers and likely imposed less cognitive load than more complex data tables typically encountered in real-world scenarios, in spite of the mental math to compute proportions.}

\begin{figure}[t]
    \centering
    \includegraphics[width=0.95\linewidth]{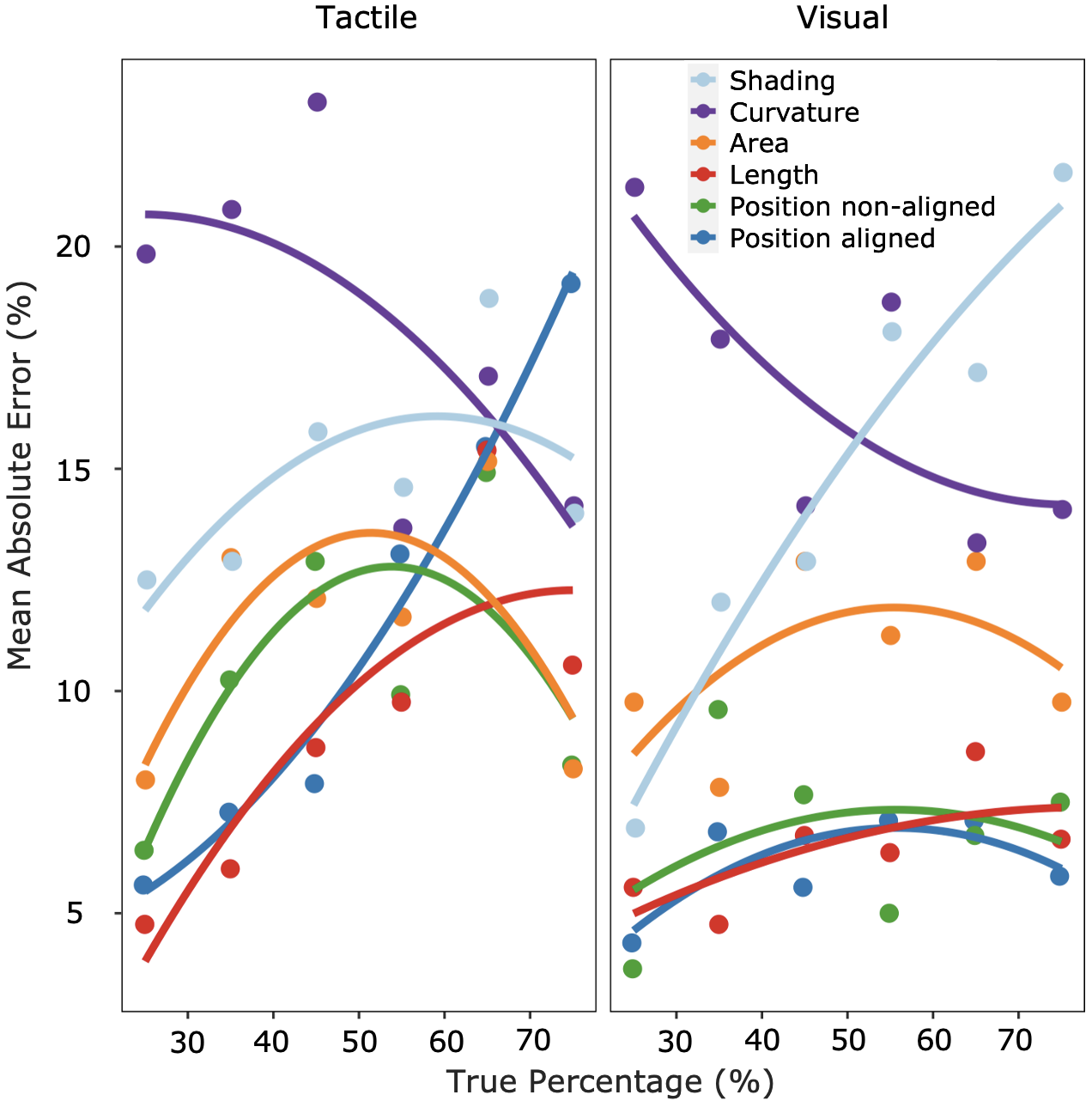}
    \caption{A side-by-side comparison of the visual and tactile modalities, where each encoding type is distinguished by color. The x-axis represents the true proportion, and the y-axis displays the absolute error (\%). \revise{The dots represent MAE associated with each true proportion. Curves are fitted using a polynomial of degree 2.} 
    }
    \vspace{-1.5em}
    \label{fig:result}
\end{figure}

\medskip
\noindent\textbf{H5: Perceived Accuracy.} 
On a scale of 1 = least accurate to 7= most accurate, participants perceived the tactile condition as less accurate than the visual condition: tactile received an average rating of 3.25 / 7 ($SD=0.75$, $range=2-4$)\remove{,} \revise{and} visual received an average of 4.5 / 7 ($SD=1.09$, $range=2-6$). \revise{A Wilcoxon signed-rank test was then performed ($V=7.5, p=0.02297$), suggesting that there is a significant difference in how participants rated their performance between the two modalities, with higher perceived accuracy in the visual condition.}

\remove{
\begin{table}[h]
    \centering
    \caption{MAE for Tactile and Visual Modalities
    \emily{posisbly replace this with the figure Kristin produced (and move table details to supplemental)}}
    \resizebox{\linewidth}{!}{
    \begin{tabular}{lcc}
        \hline
        \textbf{Encoding Type} & \textbf{Tactile Modality MAE} & \textbf{Visual Modality MAE} \\
        \hline
        Length & 9.57 & \remove{7.04}\revise{6.43} \\
        Position non-aligned & \remove{10.46}\revise{10.04} & \remove{6.71}\revise{6.45} \\
        Area & 11.36 & 10.74 \\
        Position aligned & \remove{12.71}\revise{12.04} & \remove{6.13}\revise{6.12} \\
        Shading & 14.78 & \remove{14.79}\revise{14.30} \\
        Curvature & 18.14 & 16.60 \\
        \hline
    \end{tabular}
    }
    \label{tab:mae}
\end{table}
}

\remove{\begin{figure}[t]
    \centering
    \includegraphics[width=\linewidth]{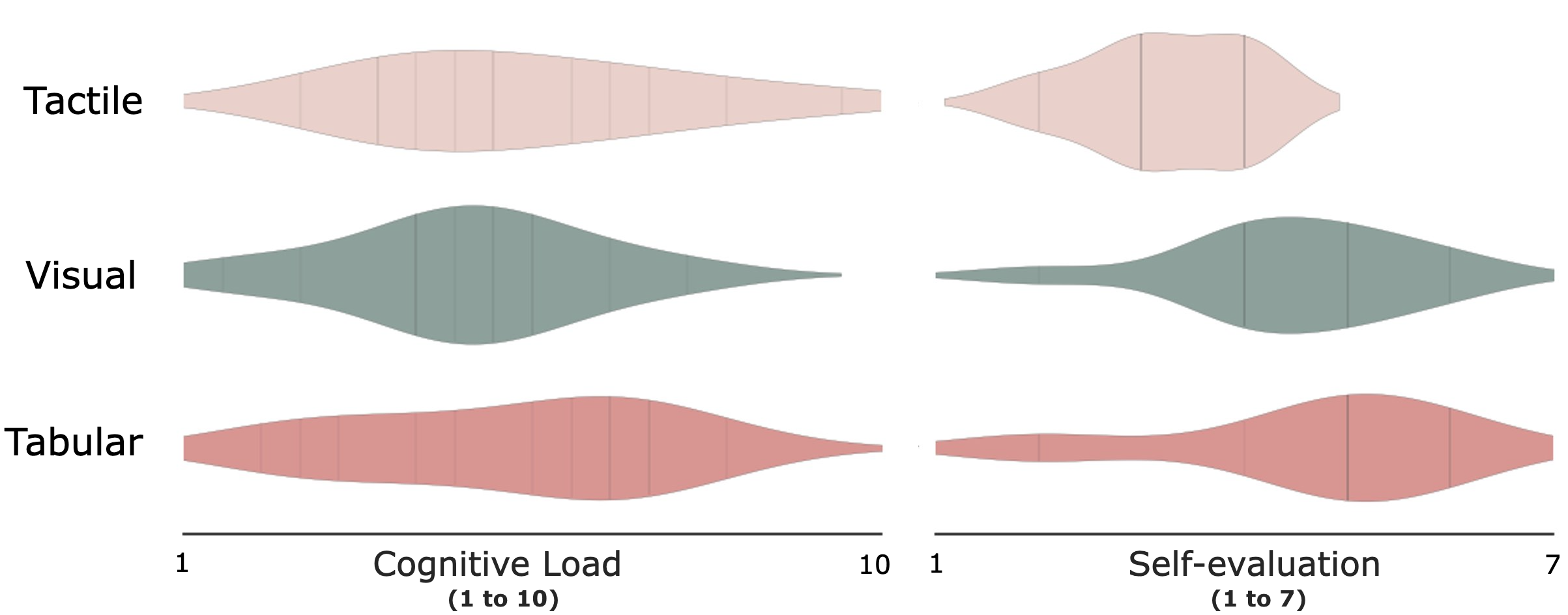}
    \caption{Perceived cognitive load and self-evaluation for each modality
    }
    \label{fig:violins}
\end{figure}}

\section{Discussion \& Conclusion}
\remove{Just as the Braille system revolutionized written communication over previous techniques of embossed letters, we echo the sentiment that there is opportunity in the visualization community to transform the way data is communicated to individuals with visual impairments \cite{lundgard2019sociotechnical}.}

In Figure ~\ref{fig:result}, we observe most encoding types had the highest error in the 50-60\% range of true percentage difference with lowest error at the extreme ends of true percentage, consistent with Cleveland and McGill~\cite{cleveland1984graphical, cleveland1986experiment}. 
Encoding types that were most accurate for visual (position aligned, position non-aligned, and length) tended to have relatively flat fitted curves, suggesting minimal changes in error as a function of true percentage. 
These same encodings for tactile (left) resulted in much higher error rates, however. 
Some encoding types can interestingly be observed to increase in error as a function of true percentage: position aligned in tactile (left) and shading in visual (right). 
When we consider raw error compared to absolute error, we further observe a systematic error of overestimation for some encoding types such as curvature and area in the visual condition, while we see underestimation for others such as shading in the visual condition and position aligned, length, and shading in the tactile condition (see Supplemental Materials). 
These observations, in conjunction with different observed rankings of perceptual accuracy, provide empirical support to prior claims that a one-to-one translation of visual encodings into a tactile form would be insufficient \cite{lundgard2019sociotechnical} and suggest the need for new guidance for tactile graphics.

\medskip
\noindent\textbf{Limitations and Future Work.} 
%
A primary limitation of our study stems from our participant group, which consisted of \revise{a small sample of 12} college students for the in-lab experiment. This demographic \revise{and sample size} likely does not encompass the full range of tactile acuity \cite{wolfe2006sensation}. 
As the field holds promise for the accessibility community, we aim to include participants with blindness or low vision in future studies to understand if differences may arise in the rankings of tactile encodings when individuals have different levels of acuity with tactile encodings, including experience with Braille.

\medskip
\noindent\textbf{Conclusion. }
In this paper, we compared perceptual accuracy for visual\remove{,} \revise{and} tactile \revise{modalities.} \remove{, and tabular modalities.} 
\revise{Our findings indicate that participants have better overall accuracy for visual modality than tactile.}
While we observed marginal differences in the order of accuracy of perception of visual encoding types (position aligned, position non-aligned, length, area, shading, curvature) found by Cleveland and McGill, we found that this ranking \textbf{deviated} for tactile encodings (length, position non-aligned, area, position aligned, shading, curvature) \cite{cleveland1984graphical, cleveland1986experiment}. 
This result highlights the current empirical gap in guiding the design of charts and graphs, and underscores the need for further foundational research in tactile data visualization.

\acknowledgments{
The authors wish to thank Matt Kay for his generous supply of clever paper titles.}

\bibliographystyle{abbrv-doi}

\bibliography{template}
\end{document}